\documentclass[A4paper,twocolumn,aps,prl,floatfix,citeautoscript,nobibnotes]{revtex4-1} 
\usepackage{epsfig,epstopdf}
\usepackage{graphicx}
\usepackage{amsfonts,amsmath,amssymb}
\usepackage{latexsym}
\usepackage{color}
\usepackage{graphicx}
\usepackage{multirow}

\begin{document}

\title{Simple microscopic model for magneto-electric coupling in type-II antiferromagnetic multiferroics}

\author{ L. Pili}
\affiliation{{ Instituto de F\'{\i}sica de L\'{\i}quidos y Sistemas Biol\'ogicos (IFLYSIB), UNLP-CONICET, La Plata, Argentina}}
\affiliation{Departamento de F\'{\i}sica, Facultad de Ciencias Exactas, Universidad Nacional de La Plata, La Plata, Argentina}

\author{ R. A. Borzi}
\affiliation{{ Instituto de F\'{\i}sica de L\'{\i}quidos y Sistemas Biol\'ogicos (IFLYSIB), UNLP-CONICET, La Plata, Argentina}}
\affiliation{Departamento de F\'{\i}sica, Facultad de Ciencias Exactas, Universidad Nacional de La Plata, La Plata, Argentina}

\author{D. C. Cabra }
\affiliation{{ Instituto de F\'{\i}sica de L\'{\i}quidos y Sistemas Biol\'ogicos (IFLYSIB), UNLP-CONICET, La Plata, Argentina}}
\affiliation{Departamento de F\'{\i}sica, Facultad de Ciencias Exactas, Universidad Nacional de La Plata, La Plata, Argentina}

\author{ S. A. Grigera}
\affiliation{{ Instituto de F\'{\i}sica de L\'{\i}quidos y Sistemas Biol\'ogicos (IFLYSIB), UNLP-CONICET, La Plata, Argentina}}
\affiliation{Departamento de F\'{\i}sica, Facultad de Ciencias Exactas, Universidad Nacional de La Plata, La Plata, Argentina}

\date{\today}

\begin{abstract}
We present a simple two-dimensional model in which the lattice degrees of freedom mediate the interactions between magnetic moments and electric dipoles. This model reproduces basic features, such as a sudden electric polarization switch-off when a magnetic field is applied and the ubiquitous dimerized distortion patterns and magnetic $\uparrow \uparrow \downarrow \downarrow$ ordering, observed in several multiferroic materials of  different composition. The list includes E-type manganites, RMnO$_3$,
nickelates such as in YNiO$_3$ and other materials under strain, such as TbMnO$_3$. In spite of its simplicity, the model presented here captures the essence of the origin of multiferroicity in a large class of type II multiferroics. 
\end{abstract}

\maketitle

\textit{Introduction.---}
Multiferroic (MF) materials, those in which electric and magnetic degrees of freedom are coupled, are a subject of growing interest, not only  for their potential technological applications but also because of the theoretical interest raised by the different unusual properties and effects discovered over the last years \cite{eerenstein2006multiferroic,hur2009giant,lee2008giant,dong2015multiferroic,fiebig2016evolution}.  Of interest to technological applications is the possibility of using multiferroicity for low-energy switching in data storage devices that could lead to future generation ultra-low-energy electronics. Coupling between the magnetic and ferroelectric orders could allow for bit-imprinting by an {\em electric} rather than  magnetic field \cite{vopson2015fundamentals,wang2016multiferroic,hu2019progress}.

Among the large family of MF materials known today, there is a special class, dubbed {\em improper type II MFs}, which are distinguished by the fact that the magnetic and ferroelectric order occur simultaneously through a cooperative transition (see e.g. refs. \onlinecite{van2008multiferroicity,khomskii2009trend,cheong2007multiferroics}). An important subclass of these materials is that in which the magnetic order is collinear at low temperatures and consists of an arrangement of spins following a period 4 pattern $\uparrow \uparrow \downarrow \downarrow$ in one, two or the three directions of the crystal, which we will refer to as \textit{uudd}.

A non-exhaustive list of materials in this special class includes: i)
A large group of nickelates, which show first a  metal--insulator transition involving a structural change, followed by a paramagnet to type-E antiferromagnetic  phase, with magnetic \textit{uudd} ordering along the three crystal directions \cite{medarde1997structural,catalano2018rare};
ii) Manganites, which are believed to exhibit both ferroelectric and antiferromagnetic transitions and in some cases, {\it e.g.} in HoMnO$_3$,  a magnetic \textit{uudd} ordering (also type-E) simultaneous with a structural change  \cite{sergienko2006ferroelectricity,dong2009double,lilienblum2015ferroelectricity};
and iii) Double-perovskites such as Yb$_2$CoMnO$_6$ \cite{blasco2015evidence},
Y$_2$CoMnO$_6$ \cite{murthy2014metamagnetic},
Lu$_2$MnCoO$_6$ \cite{chikara2016electric,zhang2016origins}, Er$_2$CoMnO$_6$ \cite{kim2019strong}, and R$_2$NiMnO$_6$ (R = Pr, Nd, Sm, Gd, Tb, Dy, Ho, and Er), where a giant magnetoelectric effect has been reported \cite{zhou2015magnetic}.

Motivated by these multiple observations we present a minimal model where a simple mechanism stabilises the ubiquitous lattice dimerization and  \textit{uudd} spin ordering. In our model, local dipoles arise from spontaneous distortions of the crystal lattice, which are in turn stabilised by the magnetoelastic coupling and affected by the consequent electric (dipole-dipole) interactions.  A related one-dimensional model proposal, that reproduces the basic phenomenology of 1D materials, has been recently analyzed in Refs. \onlinecite{cabra2019microscopic,cabra2021topological}. Related ``exchange-striction'' mechanisms to explain MF behaviour in different classes of materials have been proposed and studied (see. e.g. Ref.  \onlinecite{Jeon2009TheoryOM})

In a previous paper \cite{pili2019two} two of the authors have studied a magneto-elastic two-dimensional Ising model in which three main phases were in competition: a ferro (FM) or antiferromagnetic (AFM) phase on the undistorted square lattice, a so-called plaquette or checkerboard phase (CB), and the stripe phase (ST). The latter corresponds to an E-type \textit{uudd} magnetic ordering along the two principal crystal directions. These same two magneto-elastic phases have been also studied in the quantum spin-Peierls case \cite{Sirker2002GroundstatePO} in a square lattice. 
It was shown in Ref.~\onlinecite{pili2019two} that for phenomenologically reliable couplings the CB phase is always lower in energy. Here we add two ingredients to this purely magneto-elastic model. On the one hand, lattice deformations drive the setup of electric dipole moments via distortions of the charge environment. On the other, we take into account the dipole-dipole interactions resulting from these moments. Interestingly, we observe that this electric dipolar interactions can alter the relative values of the ground state energies of these phases, turning the ST or \textit{uudd} phase --the one relevant to the  experiments listed above-- as the stable one even within this classical framework.

We show that the magnetoelectric coupling is quite effective. Crucially, it leads to a sharp switch-off of the spontaneous polarization as a function of the applied magnetic field, in concurrence with a metamagnetic transition. These simultaneous transitions have been observed in a wide variety of materials, such as Er$_2$CoMnO$_6$\cite{kim2019strong}, Lu$_2$CoMnO$_6$\cite{yanez2011multiferroic,chikara2016electric}, and R$_2$V$_2$O$_7$ (R=Ni,Co)\cite{chen2018magnetic}. The effect is also observed in some non-collinear cases, such as TbMnO$_3$, which shows gigantic magnetoelectric and magnetocapacitance effects \cite{Kimura2003MagneticCO}. Interestingly enough, this material can be driven into an \textit{uudd} state by epitaxial strain \cite{shimamoto2017tuning}.

\begin{figure}
\includegraphics[width=\columnwidth]{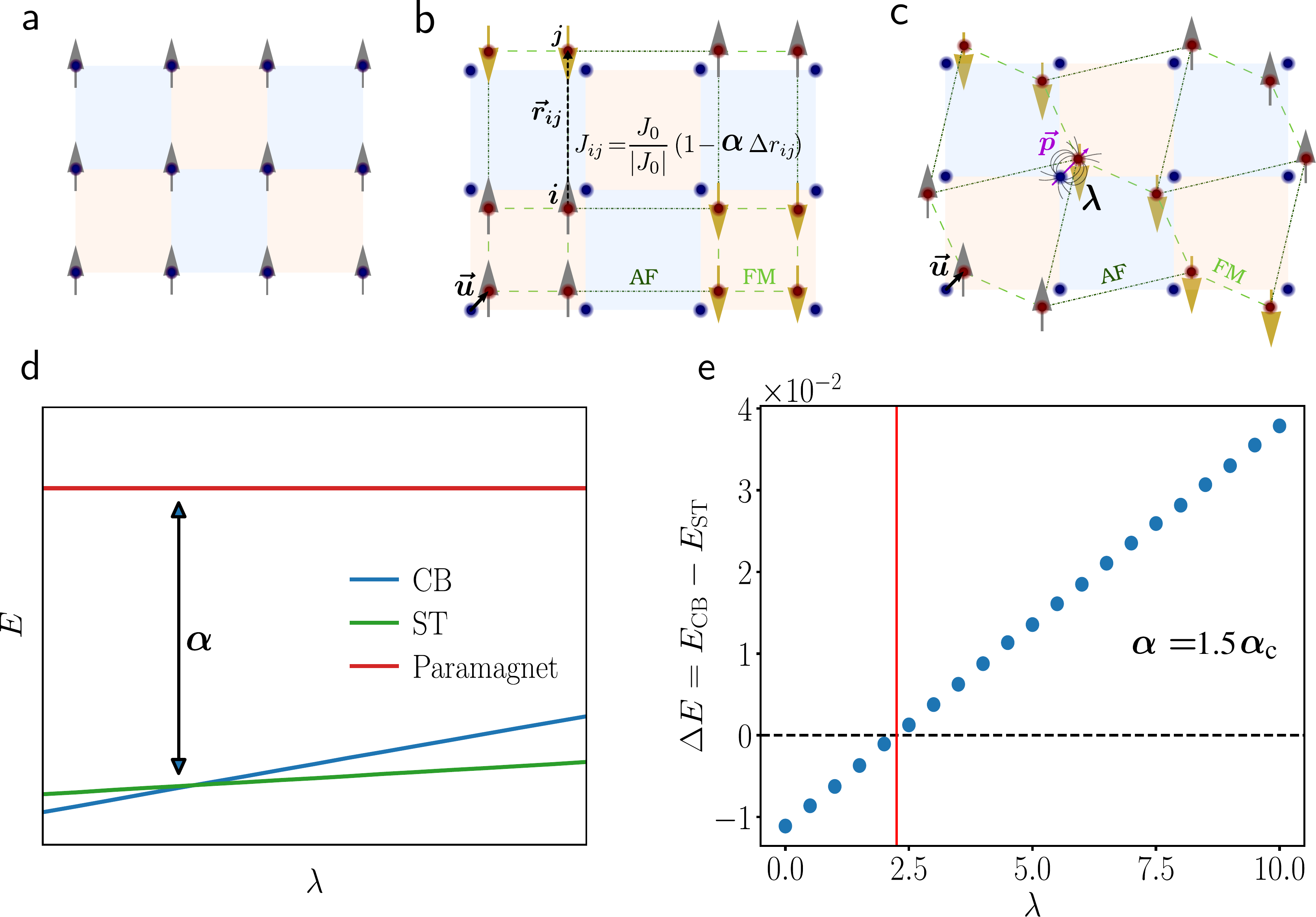}
	\caption{{\em Interaction terms and phases.} Aside from the paramagnetic phase, three ordered states are relevant to this study.  Magnetic and elastic degrees of freedom are taken on equal footing on our model, but the names chosen refer to spin configurations. a) {\em Ordered Ising state} (Ferromagnetic in the example pictured). For relatively small values of the magnetoelastic coupling $\alpha$, this is the preferred spin configuration.  Distortions average zero in this state. b) {\em Checkerboard phase} and  c) {\em Stripe phase}. If the magnetoelastic coupling constant $\alpha$ is bigger than $\alpha_c$, it is favorable to deform the lattice.  These deformations imply a dipolar electric moment $\textbf{p}_i=\gamma_i \delta \textbf{r}_i$, represented here by pairs of coloured circles. (d) Dipolar interactions, with an energy scale given by $\lambda$, do not affect the FM or the paramagnet, but change the energy balance between the ST and CB phases. (e) As shown here for the example with $\alpha = 1.5\alpha_c$, and $\gamma_i=1,2$ for odd and even sites respectively, the ST becomes the ground state of the system above $\lambda_c =2.45$.}
	\label{Fig_phases} 
\end{figure}

\textit{The Model.---}
This Ising model is based on the so-called {\em Einstein site phonon spin model} \cite{wang2008app}, 
that considers a coupling between magnetic and elastic degrees of freedom. In it, the sites have displacements given by a set of independent harmonic oscillators. This assumes that the most important lattice distortion contribution is coming from optical phonons, which is a reasonable choice given that in real materials the active magnetic lattice is usually a sub-lattice of a more complex crystal structure. The model presented here incorporates electrical properties by assuming that each site displacement implies the formation of a dipolar electric moment; dipole-dipole interactions are then considered up to first nearest neighbours. The Hamiltonian is given by:
\begin{multline}
\frac{\mathcal{H}}{|J_0|}=\sum_{\langle i,j \rangle} J(r_{ij}) S_i S_j+\frac{K_e}{2} \sum_i (\delta r_{i})^2 + \\
\lambda \sum_{\langle i,j \rangle} [\textbf{p}_i \cdot \textbf{p}_j -3 (\textbf{r}_{ij}^0 \cdot \textbf{p}_j)(\textbf{r}_{ij}^0 \cdot \textbf{p}_i)]+\mathcal{H}_{\text{field}}    
\end{multline}
\noindent where $S_i$ stands for an Ising type spin pointing along the [0 1] crystal direction ($S_i=\pm 1$) at position $\textbf{r}_i$ measured in units of the cell constant. The $\textbf{r}_{ij}=\textbf{r}_i-\textbf{r}_j$ are the relative position vectors of different spins. We call $\delta \textbf{r}_i \equiv \textbf{r}_i-\textbf{r}_i^0$ the site displacement, and $\delta r_{ij} =|\textbf{r}_{ij} -\textbf{r}_{ij}^0|$ the distance change between sites (see Figs.~\ref{Fig_phases}b and c). The distance-dependent\footnote{This dependence can also be thought of as a parametrization of an angle dependent interaction, when it is mediated by a non-magnetic ion (as in HoMnO$_3$).} exchange constant is given by
\begin{equation}
J(r_{ij}) = \text{sgn}(J_0) \ [1-\alpha\delta r_{ij}].
\end{equation}
Here $\alpha$ is the magnetoelastic coupling, while the electric dipole moment at site $i$ is given by $\textbf{p}_i = \gamma_i \delta \textbf{r}_i$.  The proportionality constant, $\gamma_i$, can be site dependent if the system is composed of sublattices with different ionic charges.

External magnetic and electric fields are taken into account by the term:
\begin{equation}
\mathcal{H}_{\text{field}}= -B \sum_i S_i - \sum_i \textbf{E} \cdot \textbf{p}_i,
\end{equation}
where $B$ points trivially along the $y$ direction and $E$ points along the diagonal directions of the lattice.

To simulate the elastic distortions we consider polar coordinates  $(\rho,\theta)$ to describe each $\delta r_i$.  The angle $\theta$ is treated like in a clock model of $360$ equally spaced angles,  and the displacement $\rho$ is chosen randomly in a distribution from $0$ to a temperature dependent maximum $\delta_{max}(T)$. The use of the latter has no impact on the results obtained from the simulation; it is introduced as a way to optimise speed by avoiding the proposal of extremely unlikely moves at low temperatures \cite{pili2019two}.

In accordance with the spirit of the model, the magnetic and elastic degrees of freedom are treated simultaneously in the Monte Carlo simulations. We assume that the relaxation times of the magnetic degrees of freedom are much shorter than the elastic ones. Each step of the simulation is split into elastic and magnetic moves. Our assumption, similar to the Born-Oppenheimer approximation, translates into the fact that each elastic move is done with a relaxed magnetic configuration~\cite{pili2019two}. 

\textit{Results for $\gamma_i = 0$: purely magnetoelastic system.---} 
The magneto-elastic phase diagram of the model in the absence of any polarization effects has been studied in Ref. \onlinecite{pili2019two}.  As the magneto-elastic coupling, $\alpha$, is increased, the critical temperature of the ordered FM or AFM Ising phase  decreases steadily, reaching $T=0$ at a critical coupling $\alpha_c=\sqrt{K_e/2}$, above which the system goes through a simultaneous magnetic and structural transition.  The ground state becomes a {\em checkerboard} of ferromagnetic clusters, aligned antiferromagnetically (see Fig. \ref{Fig_phases}b)).  The critical temperature of the CB phase increases with increasing $\alpha$.  For values of $\alpha$ slightly above $\alpha_c$ there is another phase with  long-range {\em uudd} order,  the ST state, with energy comparable to the ground state.  This state, pictured in Fig.~\ref{Fig_phases}c), consists of diagonal ferromagnetic stripes aligned antiferromagnetically.  While the CB state is the ground state, the energetic proximity of the stripe state means that additional interactions might easily reverse the situation.  As we will see, this is the effect of electric dipolar interactions.

\begin{figure}
\includegraphics[width=0.9\columnwidth]{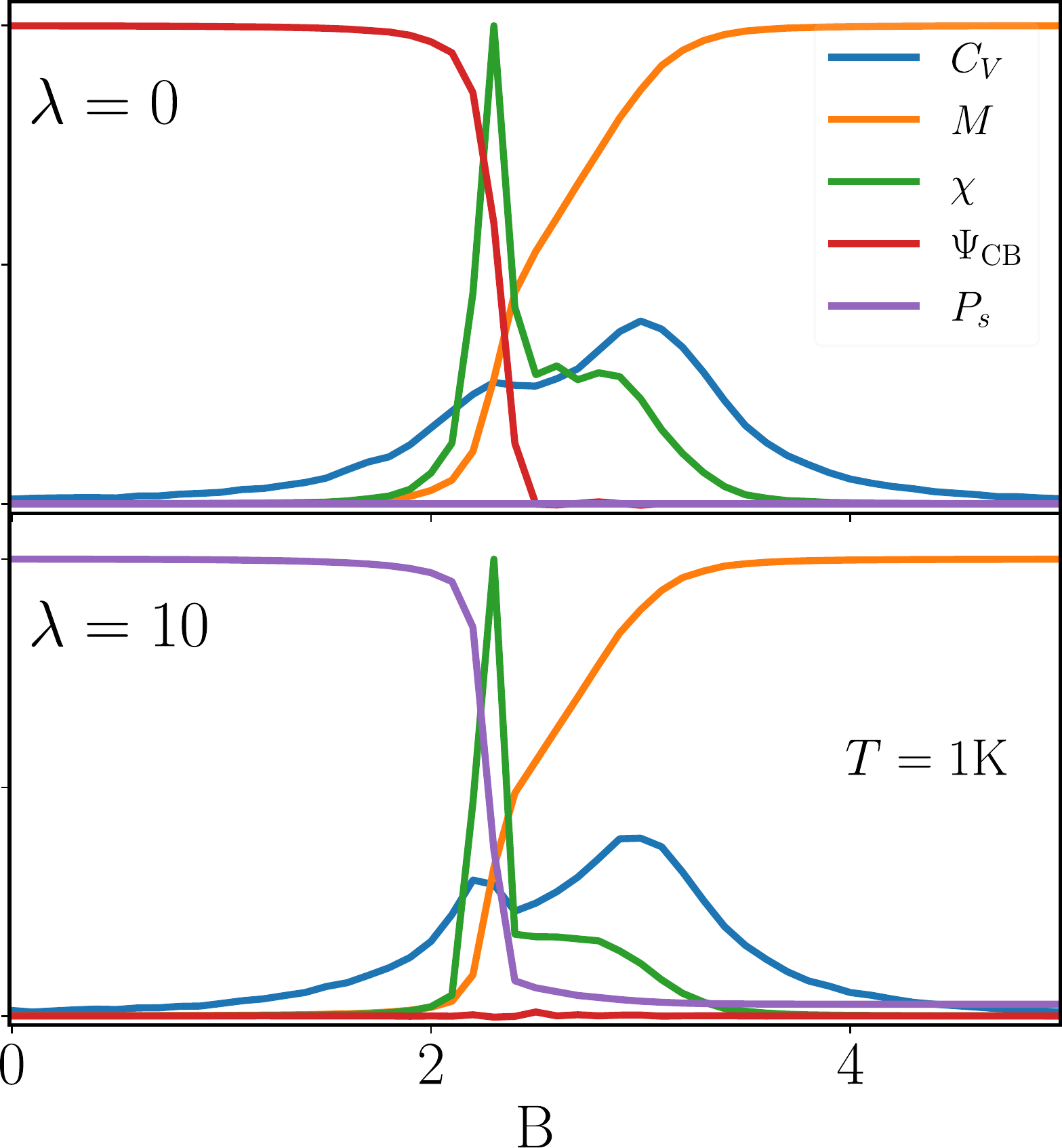}
	\caption{{\em Effect of the magnetic field on charge and magnetic degrees of freedom (case of homogeneous charge distribution $\gamma_i=\gamma$)}.  Susceptibility peaks, and the first peak in $C_V$ coincide with the abrupt fall of $P$ and the rise in $M$. The behaviour is similar for both $\lambda$ (i.e., for both ground states). In both cases $\alpha=1.5\alpha_c$.   The rise in $M$ is smooth and goes through an intermediate step marked by peaks in the specific heat.}
	\label{Fig_equal} 
\end{figure}
%

\textit{Results for $\gamma_i \ne 0$: multiferroicity .---} 
When $\gamma_i \ne 0$, electric dipole moments are developed that are proportional to the local site displacements. We begin by analysing the simplest (homogeneous) case, i.e. $\gamma_i=\gamma$ for all $i$.  In the paramagnetic or in the FM/AFM ordered phases this is irrelevant: their minimum energy is achieved without distorting the square lattice.  This is no longer true for  $\alpha > \alpha_c$, and, as it can be seen in Figs.~\ref{Fig_phases}b and \ref{Fig_phases}c, both the checkerboard and the stripe states develop electric moments at every site, albeit with different configurations.  When the dipolar interaction between these moments, proportional to $\lambda$, is taken into account to first nearest neighbours, the balance between the energies of the CB and ST states changes.  This is plotted  in  Fig.~\ref{Fig_phases} d for a fixed value of $\alpha > \alpha_c$.  While the paramagnetic phase is trivially unaffected, the energy of both the CB and the ST phases grows linearly, with different slopes for each case. As shown in Fig.~\ref{Fig_phases} d, there is a critical $\lambda_c$ above which the ST phase becomes the ground state of the system.

For $\lambda > \lambda_c$ the ground state is a stripe phase.  If $\gamma_i \ne 0$ is equal for all $i$, the order in the ST phase is antiferroelectric, since the sum of the displacements cancels out. 
The magnetic and electric dipole directions are not correlated with each other: the orientation of one does not determine the other. 

\begin{figure}
\includegraphics[width=0.9\columnwidth]{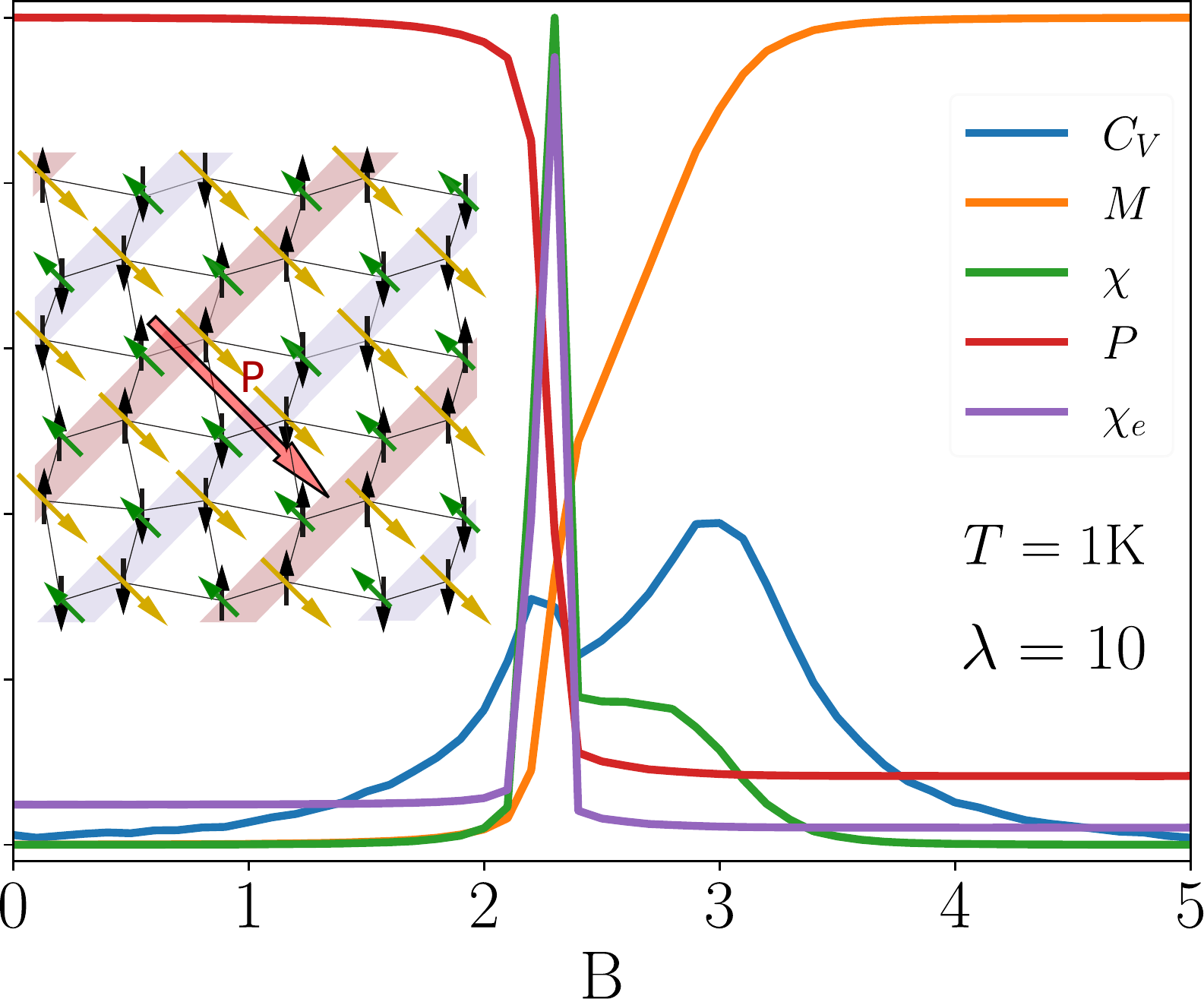}
\caption{{\em Effect of the magnetic field on charge and magnetic degrees of freedom (non-homogeneous charge distribution $\gamma_i$).} In this case (sketched in the inset) where there are two sublattices with different charges the system has a non-zero homogeneous polarization $P$ at low fields. Like in the case with $\gamma_i=\gamma$, the transition towards the high-field state is also marked by an intermediate phase. 
}
\label{Fig_multi} 
\end{figure}

There is a simple and physically sensible way in which a non-zero bulk polarization can arise in this model.  A common type of crystal is composed of two interpenetrated square sublattices of different ions.  If the polarizability of these sites is different, which can be easily taken into account in the model by making $\gamma_i$ different in each sublattice, the ground state is not affected (see  Fig. \ref{Fig_phases} e) and there is a net polarization of the whole system (see inset of Fig.~\ref{Fig_multi}).

\begin{figure}
\includegraphics[width=1.\columnwidth]{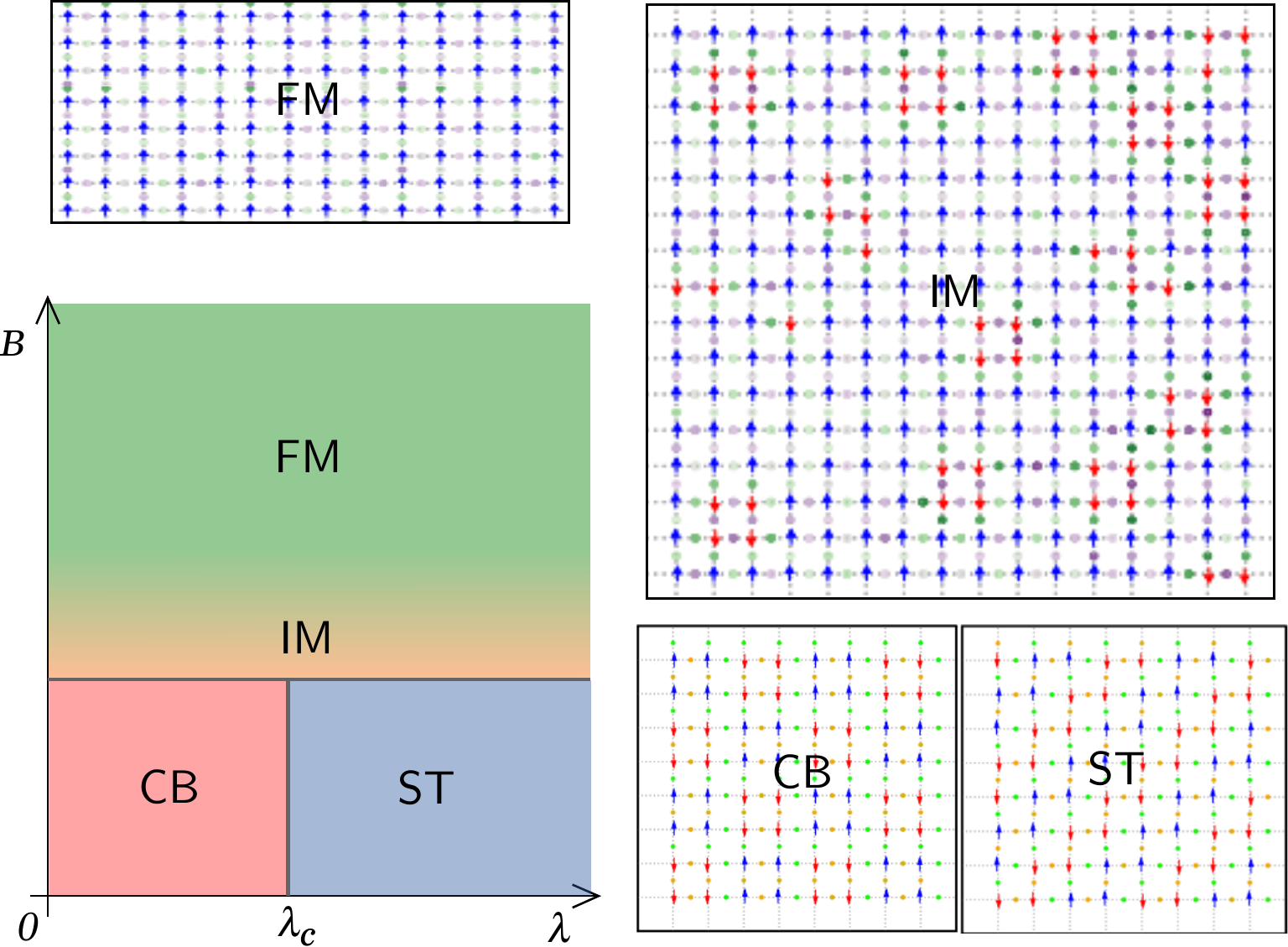}
	\caption{{\em $B-\lambda$ phase diagram.}  For low fields the ground state of the system transitions from a checkerboard (CB), below $\lambda_c$, to a stripe phase (ST).  As the field is increased, both the CB and ST states transition into a ferromagnet (FM) going through an intermediate state(IM) where the system is mostly ordered ferromagnetically, with some remnants of the low-field phase in the form of aligned squares that oppose the magnetic field direction. }
	\label{PhD} 
\end{figure}
%

\textit{The effect of an external magnetic field.---}
For $\alpha > \alpha_c$, both above and below $\lambda_c$, the magnetic ground states are different forms of antiferromagnetism. An external magnetic field eventually drives the system into a homogeneous FM state. Fig. \ref{Fig_equal} shows the behaviour of the magnetisation, magnetic susceptibility, and the specific heat as a function of the externally applied magnetic field both below $\lambda_c$ (top panel, $\lambda =0$) and above $\lambda_c$ (bottom panel, $\lambda =10$).  In the first case, the magnetic ground state is a  CB and its order parameter $\Psi_{\rm CB}$ (see appendix) is also shown.  In the second case, where the ground state is a ST, it is the staggered polarization, $P_s$, that is plotted as a function of the field.   As shown in the figure, the behaviour of the system is very similar regardless the ground state: the antiferromagnetic state with $M=0$ is preserved for low fields and eventually gives way to the FM state through a metamagnetic transition.  The field at which this transition starts is very similar for both cases, which is to be expected, given the subtle energy difference between both magnetic ground states. 
There is a structure at the transition, evidenced as a double peak in the specific heat, and as a peak and shoulder in the magnetic susceptibility

The $B-\lambda$ phase diagram can be simply sketched (Fig.~\ref{PhD}): at low fields there is a transition between the CB and ST states as a function of $\lambda$.  As the field is increased, the system polarizes into a FM state going through an intermediate state (IM), the existence of which is marked by the two peaks in the specific heat.  Snapshots of the different phases, also shown in Fig.~\ref{PhD}, give further information about the intermediate state:  here the system is mostly ordered ferromagnetically, but retains some remnants of the low-field phases in the form of aligned clusters that oppose the magnetic field direction.

The effect of introducing a non-homogeneous $\gamma_i$ by making $\gamma_i$ different in two sublattices (as sketched in the inset of Fig.~\ref{Fig_multi}) leaves unchanged the overall behaviour of the system as a function of magnetic field (Fig.~\ref{Fig_multi}).  The crucial difference, in terms of experimental observables, is that in this case the system switches from a homogeneous $P \ne 0 $ and $M=0$ at low fields, to a negligible polarization $P \approx 0$ and a saturated $M \ne 0$ at high fields. The peaks in $\chi_M=dM/dB$ and $\chi_e=dP/dB$ coincide almost exactly.  In this way the model reproduces the polarization switch-off observed in many experimental systems\cite{kim2019strong,yanez2011multiferroic,chikara2016electric,chen2018magnetic,Kimura2003MagneticCO}. 

%

\textit{Scattering signatures.---}
The ST and CB phases described before have signatures in scattering experiments, both in neutron scattering (the spin-channel), coming from the different long-range ordered magnetic structures, and in X-ray scattering (the charge channel) as a consequence of their characteristic distortion patterns.  These experimentally accessible characteristics can be calculated from the MC simulations (see Sup. Info). 

\begin{figure}
\vspace{2ex}
\centerline{\sf Spin channel \hspace{17ex} Charge channel}
\vspace{1ex}
\includegraphics[width=0.47\columnwidth]{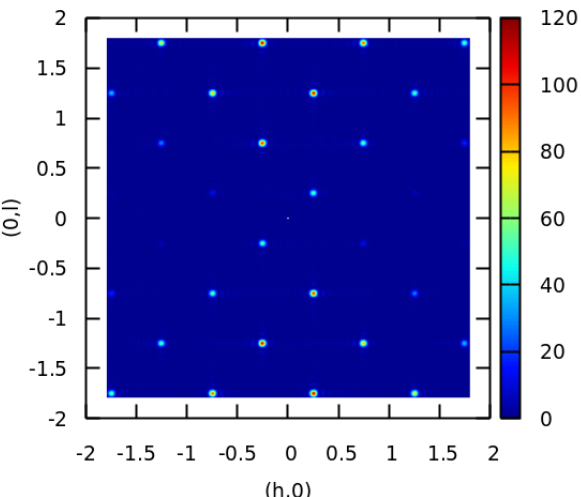} ~~\includegraphics[width=0.47\columnwidth]{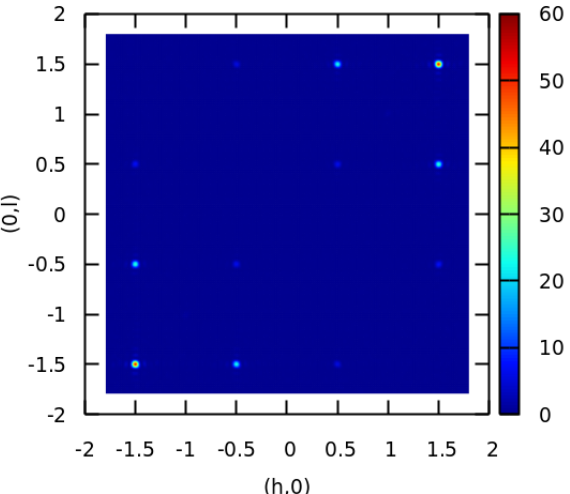}
\includegraphics[width=0.47\columnwidth]{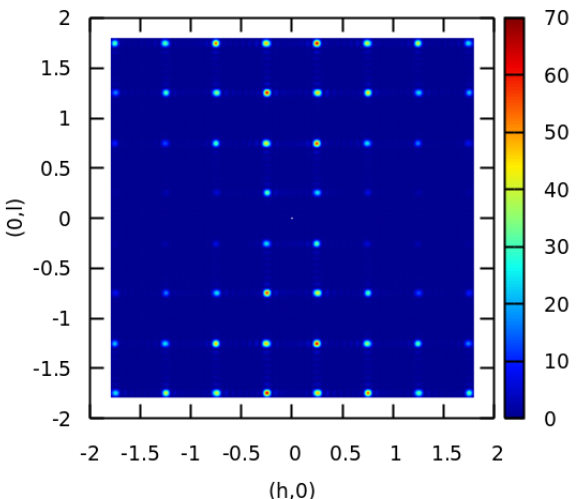} ~~\includegraphics[width=0.47\columnwidth]{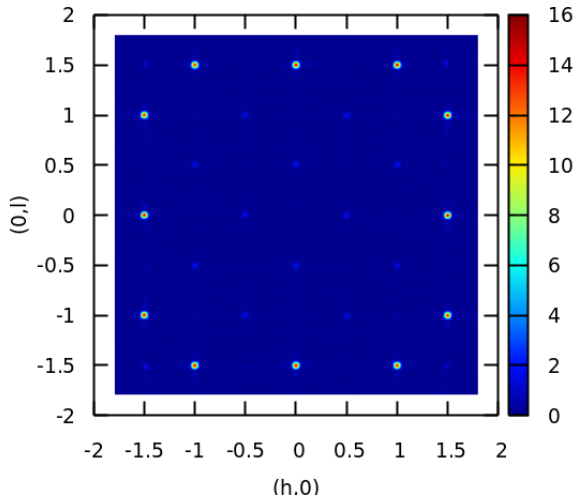}
	\caption{{\em Signatures in scattering experiments.} Structure factors in the spin  channel (left) and diffuse charge channel, associated with the displaced charges (right).  The stripe phase (up) shows C2 symmetry in both channels, while the checkerboard phase (down) retains the C4 symmetry of the lattice.}
	\label{Fig_neutrons} 
\end{figure}

Figure \ref{Fig_neutrons} shows the structure factors in the spin  channel (left) and diffuse charge channel (right) for the ST phase (up) and the CB phase (down).  As expected, for both channels, the stripe phase shows C$_2$ symmetry, while the checkerboard phase  retains the C$_4$ symmetry of the lattice (albeit with a different unit cell).

\textit{Summary and Conclusions.---}
In this work we present what is probably the simplest possible model that reproduces the basic features observed in several multiferroic materials, such as E-type manganites, RMnO$_3$, nickelates such as in YNiO$_3$, double-perovskites and other materials under strain, such as TbMnO$_3$.  The model, based on the Einstein site phonon spin model, is a nearest neighbour Ising model on a square lattice that adds coupling between magnetic and elastic degrees of freedom.  The latter has two effects: first it alters the local exchange interaction, and second, it gives rise to electric dipole moments, which in turn interact with a NN dipolar term.  The presence of these two interactions changes the ground state of the system from the usual FM or AFM ordered state (depending on the sign of $J$) to a striped state where FM and AFM couplings coexist, with magnetic \textit{uudd} ordering,  and where a non-zero polarization can develop.  An applied magnetic field eventually switches-off the electric polarization, driving the system into an ordered FM state through a metamagnetic transition.

The model presented here captures the essence of the origin of multiferroicity in a large class of type II multiferroics and given its simplicity is a promising toy model to further investigate these phenomena. Understanding the role of lattice distortions in the magnetoelastic coupling would also provide a useful guide to experiments under tensile strain and film depositions on mismatched substrates.

\begin{acknowledgements}
\textit{Acknowledgements.---}
We would like to acknowledge discussions with M. Medarde and financial 
support from the Agencia Nacional de Promoci\'on Cient\'\i fica y Tecnol\'ogica (ANPCyT) through PICTs 2017-2347 and 2015-0813 and from the Consejo Nacional de Investigaciones Cient\'\i ficas y T\'ecnicas (CONICET) through PIP 0446.
\end{acknowledgements}

%

\newpage
\clearpage

\setcounter{equation}{0}
\setcounter{figure}{0}
\setcounter{page}{1}
\thispagestyle{empty} 
\makeatletter 
\renewcommand{\thefigure}{S\arabic{figure}}
\renewcommand{\theequation}{S\arabic{equation}}
\setlength\parindent{10pt}

\onecolumngrid

\begin{center}
	{\fontsize{12}{12}\selectfont
		\textbf{Supplementary Information for\\Simple microscopic model for magneto-elastic coupling in type-E antiferromagnetic multiferroics\\[5mm]}}
	{\normalsize L. Pili, R. A. Borzi, D. C. Cabra, S. A. Grigera \\[1mm]}
\end{center}
\normalsize

The supplementary material is composed of two sections:
\begin{enumerate}
    \item[I)] Definition of the CB order parameter
    \item [II)] Calculation of the neutron structure factor
\end{enumerate}

\section{I) Definition of the CB order parameter}

\begin{figure}[h]

\includegraphics[width=0.25\columnwidth]{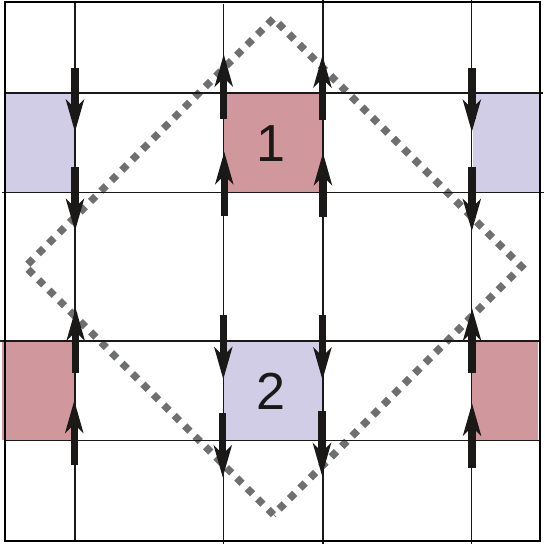} 
\caption{Unit cell and labels used for the calculation of $\Phi_{\rm CB}$  }
	\label{Fig_UC} 
\end{figure}

We use the order parameter for the checkerboard phase as defined in Ref. \onlinecite{pili2019two}.  For this we use a unit-cell like the one shown in figure \ref{Fig_UC}. The index $j$ is defined that it runs over all squares in the lattice counting as odd and even the squares marked with 1 and 2 respectively in the picture, and an index $a$ that runs over the spins in the squares. There are four possible degeneracies of the ground state (plus time reversal), corresponding to where the coloured squares are set in the unit cell.  We then define an order parameter $\Phi_{\rm CB}$ that is the sum over the four possibilities,
  \begin{equation}
 \Phi_{\rm CB} = 1/N\sum_{m=0}^4 (-1)^m|\Phi_m|,
\end{equation}
 where
\begin{equation}
\Phi_m =   \sum_{j=1}^{N/4} \sum_{a=1}^4 (-1)^j e^{i\phi_a^m}\sigma_a^j.
\end{equation}
Here $\sigma_a^j$ are Ising-spin variables that can take the values $\pm 1$, $N$ is the total number of spins, and the $\phi_a^m$ are the phase factors for the spin that take into account the four possible degeneracies: $\phi^1 = \pi (0, 0, 0, 0 ),\phi^2 = \pi (1, 0, 1, 0 ),\phi^3 = \pi (1, 1, 0,0 ),\phi^4 = \pi (1, 0, 0, 1 )$.

%

\section{II) Calculation of the neutron structure factor}

The simulated neutron structure factors have been calculated following the expression:

\begin{equation*}
    I^{Spin}(\boldsymbol{k})=\frac{1}{N}\sum_{ij}\langle S_i S_j\rangle\,\left(\boldsymbol{\mu}^{\perp}_i\cdot\boldsymbol{\mu}^{\perp}_j\right)\,e^{i\boldsymbol{k}\cdot \boldsymbol{r}_{ij}}
\end{equation*}

\noindent where $i$ and $j$ sweep the square lattice, $N$ is the number of spins, and $\langle ... \rangle$ represents thermal average (in this case, that of the product of spins at sites $i,j$). The spin quantization directions are given by $\{ \hat{\mu}_i \}$ (parallel to the $\langle 01 \rangle$ directions). Then, $\boldsymbol{\mu}^{\perp}_i$ is the component of $\hat{\mu}_i$ of the spin $\boldsymbol{S_i} = S_i \hat{\mu_i}$ at site $i$ perpendicular to the scattering wave vector $\boldsymbol{k}$: 

\begin{equation}
    \boldsymbol{\mu}^{\perp}_i=\hat{\mu}_i-\left(\hat{\mu}_i\cdot\frac{\boldsymbol{k}}{|\boldsymbol{k}|}\right)  \frac{\boldsymbol{k}}{|\boldsymbol{k}|}.
\end{equation}

For the  \textit{diffuse} structure factor associated to the displaced ions, assuming an atomic form factor unity, we calculated: 

\begin{eqnarray}
    I^{el.-dip.}(\boldsymbol{k})=\frac{2}{N}\sum_{\alpha\beta} &&\langle (e^{i\boldsymbol{k}\cdot \delta \textbf{r}_{\alpha}}-q^{av}(\boldsymbol{k}))  (e^{-i\boldsymbol{k}\cdot \delta \textbf{r}_{\beta}}-q^{av}(\boldsymbol{k}))\rangle  e^{i\boldsymbol{k}\cdot \boldsymbol{r}_{\alpha \beta}},\nonumber
\end{eqnarray}

\noindent where $q^{av}(\boldsymbol{k})=\langle e^{i\boldsymbol{k}\cdot \delta \textbf{r}_{\alpha}} \rangle$ is an average, $k-$dependent ``charge" in the perfect square lattice. 

In both cases we have thermally averaged over sets composed of $500-1000$ independent configurations for a system size $L=16$.

\end{document}